\documentclass[doublespacing]{elsart}

\usepackage{epsfig}

\begin{document}

\begin{frontmatter}



\title{Coherent bremsstrahlung and GDR width from $^{252}$Cf cold fission}


\author[label1]{Deepak Pandit},
\author[label1]{S. Mukhopadhyay},
\author[label2]{Srijit Bhattacharya},
\author[label1]{Surajit Pal},
\author[label3]{A. De},
\author[label1]{S. R. Banerjee\corauthref{cor}}
\corauth[cor]{Corresponding author.}
\ead{srb@veccal.ernet.in}


\address[label1]{Variable Energy Cyclotron Centre, 1/AF-Bidhannagar, Kolkata-700064, India}
\address[label2]{Department of Physics, Darjeeling Government. College, Darjeeling-734101, India}
\address[label3]{Department of Physics, Raniganj Girls' College, Raniganj - 713347, India}

\begin{abstract}
The energy spectrum of the high energy $\gamma$-rays in coincidence with the prompt $\gamma$-rays has been 
measured for the spontaneous fission of $^{252}$Cf. The nucleus-nucleus coherent bremsstrahlung of the 
accelerating fission fragments is observed and the result has been substantiated with a theoretical 
calculation based on the Coulomb acceleration model. The width of the giant dipole resonance (GDR) decay from the excited fission fragments has been extracted for the first time and compared with the thermal shape fluctuation model (TSFM) in the liquid drop formalism. The extracted GDR width is significantly smaller than the predictions of TSFM.  
\end{abstract}

\end{frontmatter}


Nuclear fission has been a subject of incessant research for decades. This nuclear phenomenon can occur spontaneously 
as a natural decay process or can be induced through the absorption of a relatively low-energy particle, 
such as a neutron or a photon. Since large amount of energy is available, the spontaneous fission of $^{252}$Cf has prompted various searches, in particular, for bremsstrahlung emission \cite{kas89, luke91, ploeg92}, 
neutral pions and charged pions \cite{cer88, stan89} and various exotic radioactivities \cite{ion86}. 
Photon has evoked an extra attention over pion since it is not seriously affected by absorption phenomenon in the medium. Hence, it can serve as an excellent probe to study the reaction dynamics in 
the early stage of the reaction. An accurate measurement of $\gamma$-ray emission 
from spontaneous fission reaction could throw some lights on nuclear dissipation at 
low temperature. However, in recent decades, a few experiments have been performed to explore the 
high energy part of the $\gamma$-spectrum coming from spontaneous fission. In addition to that, the observations have been found contradictory in nature.

The $\gamma$-ray energy spectrum, above 20 MeV, emitted in the spontaneous fission of 
$^{252}$Cf has been one of the fundamental problems of nuclear fission physics. In a few experiments in the past 
\cite{luke91, die74}, the yield of  $\gamma$-rays at such high energy could not be detected, while in three other  experiments, the energy spectrum could be measured \cite{kas89, ploeg92, var05}. The photons with energy 20-120 MeV are associated with the coherent bremsstrahlung of the fission fragments in the Coulomb field. In recent past,
detailed macroscopic calculations of the bremsstrahlung yield from the spontaneous fission source has been performed 
considering different acceleration models (instantaneous, pure Coulomb) \cite{kas89,luke91}, including fragment-fragment barrier penetration (tunneling)\cite{luke91}. But the high energy photon spectra extracted from those different theoretical models differ by several orders of magnitude. The conflicting experimental results as well as the theoretical calculations motivate one to carry out further investigation. 

The low energy (8-20 MeV) part of the photon spectrum is mainly associated with direct excitation 
of the giant dipole resonance (GDR) from the daughter nuclei arising in the fission process. 
An interesting feature of spontaneous fission is that the fragments are produced at low excitation
energies. At these energies, the GDR emission will only be from the first decay step \cite{hara01, gaard01, snov01}. 
As a result, the spontaneous $^{252}$Cf source provides us an unique tool to study the GDR width at low temperature (T)  and angular momentum (J), which has been a perplexing topic in recent years.
The exploration of the GDR width in fusion evaporation reaction is a 
complex process as it requires the decoupling of the effects of both J and T on the GDR width.  
In spontaneous fission, since the angular momentum is very small ($\sim$ 6$\hbar$), the GDR width will 
only be affected by temperature. 

In this letter, we report on an extensive investigation of high 
energy $\gamma$-ray yield from the coherent nucleus-nucleus bremsstrahlung  as well as the decay of  
GDR from the $^{252}$Cf cold fission. The width of the GDR decay from the excited fission fragments has been 
extracted for the first time and compared with the Thermal Shape Fluctuation Model (TSFM) \cite{kus98}.

High energy gamma-rays from the spontaneous fission of $^{252}$Cf (3$\mu$Ci) were detected in coincidence 
with the low energy discrete $\gamma$-rays emitted from the decay of excited fission fragments 
in order to establish a correlation (photons/fission) between the high energy $\gamma$-rays and 
the fission process. The source was placed as close as possible to the four multiplicity detectors \cite{dipu}, arranged in a 2 $\times$ 2 matrix, to get a start trigger in order to separate/reject the neutrons and cosmic pile ups. 
The high energy $\gamma$-rays were measured using the array LAMBDA  \cite{supm}. The array was assembled  in a 7 $\times$ 7 matrix and kept at a distance of 35 cm from the $^{252}$Cf  source. A master trigger was generated by taking a coincidence between the start trigger and any one of the 49 detectors in the pack above a high threshold of 4 MeV ensuring the
selection of fission events and rejection of background. Time of flight measurement distinguished 
the gamma-rays from neutrons while long/short gate technique was applied to reject the pile up 
events. Data were collected in this $\gamma$ - $\gamma$ coincidence mode for 450 hours. At the photon 
energies E$_\gamma$ $\geq$ 25 MeV, cosmic ray showers are the major source of background. Therefore extreme 
precaution was taken to suppress the background and obtain the experimental data free 
from cosmic impurity. Lead bricks were used as a passive protection shield from cosmic gamma-rays. 
Large area plastic scintillator pads (paddle) were used as active shielding that surrounded the LAMBDA array as well 
as the multiplicity filter to reject the cosmic muons. 
Further, the cosmic pile up events were rejected using our cluster summing technique \cite{srij} in which the 
energy deposit in each element was required to satisfy the adequate gating employed by the pulse 
shape discrimination gate and the sharp prompt time gate. Finally, the random coincidence events 
were rejected by subtracting the background spectrum which was also collected for 450 hours without 
the fission source in an identical configuration.

\begin{figure}
\begin{center}
\includegraphics[height=7.5 cm, width=7 cm]{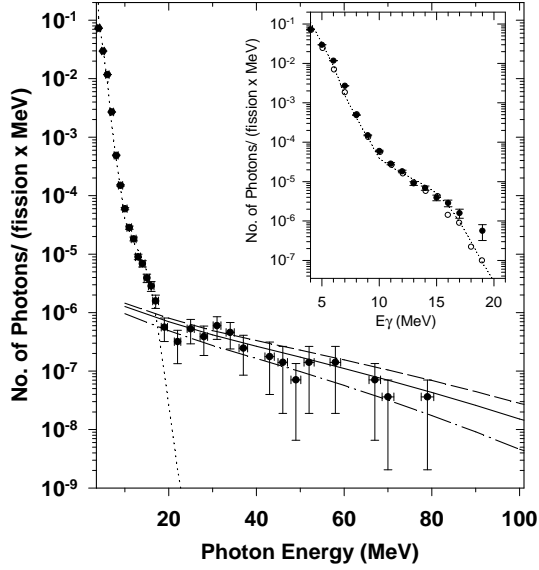}
\caption{\label{brem}The experimentally measured $\gamma$-spectrum (filled circles). The dotted line represents the CASCADE calculation.The solid, dot-dashed and dashed lines are explained in the text.
[Inset] The experimental data compared
with a previous result \cite{ploeg95}(open circle).}
\end{center}
\end{figure}

The high energy $\gamma$ spectrum measured upto 80 MeV is shown in Fig.\ref{brem} 
(filled circles). The data is compared with the neutron corrected data
(open circle, inset of Fig.1) obtained earlier in fission-$\gamma$ coincidence experiment \cite{ploeg95}.
The slope of the $\gamma$ spectrum changes sharply after 20 MeV, clearly indicating that the mechanism of the emission of photons below and above 20 MeV are completely different in origin. A theoretical calculation based on the Coulomb acceleration model in accordance with the work done by Luke et. al. \cite{luke91} was performed to estimate the photon yield above 20 MeV. 

\begin{eqnarray}\label{cou} 
\frac{d^{2}N}{dE_{\gamma}d\Omega}& = & \frac{\mu^{2}}{4\pi^{2}({\hbar}c)c^{2}}\frac{e^{2}}{E_{\gamma}} \cdot
\nonumber \\ 
& &   \left|\int dt \left[ \hat{n}\times\ddot{x}\right]e^{-i\omega t} 
\times \left(\frac{z_{1}}{m_{1}}e^{i(w/c)(\mu/m_{1})\hat{n}.x}  -\frac{z_{2}}{m_{2}}e^{-i(w/c)
(\mu/m_{2})\hat{n}.x}\right) \right|^{2} \nonumber \\
& &
\end{eqnarray}

Equation \ref{cou} gives the exact energy spectrum, in the classical nonrelativistic limit, of the bremsstrahlung produced from the acceleration of the two charged fission fragments \cite{luke91}. 
In ref\cite{pap98}, the analogous case of alpha emission was discussed in detail. Here we take only an approximate approach.
In order to solve the above equation, time (t) was expressed as a function of the distance (x) between the two fragments. The motion of the fragments was determined by solving the differential equation for the two particles under the influence of a repulsive Coulomb potential 

\begin{equation}\label{con}
\frac{1}{2} \mu \dot{x}^2 +\frac{k}{x} = E
\end{equation}

where k is Z$_1Z_2e^2$ and E is the total energy of the system. The expression for t(x) was calculated from eqn. \ref{con} and  substituted in eqn. \ref{cou}. Next, the integral was carried out numerically in position space. 
The minimal distance between the two fission fragments (x$_{scission}$=Z$_1Z_2e^2$/E) is critical and determines strongly the yield of the bremsstrahlung. 
For the most probable fission pair (A = 109, Z = 43 and A = 143, Z = 55), the experimentally measured total kinetic energy is 187 MeV \cite{sch66} which yields a value of 18.2 fm for x$_{scission}$ assuming no pre-scission kinetic energy. Due to the tunneling process, the actual acceleration starts before scission and at the scission point the kinetic energy is about 25-30 MeV.  The corresponding values of x$_{scission}$ is 21.0 and 21.7 fm, respectively \cite{ploeg}. The classical bremsstrahlung takes over from that point. The calculation was performed for a distribution of the most probable masses and charges arising from the fission of $^{252}$Cf \cite{wahl88} both including and excluding the pre-scission kinetic energy. The photon spectrum  was averaged over the total solid angle considering an isotropic emission since the angular correlation between the fission fragments and the high energy $\gamma$-rays could not be ascertained in this case. Finally, the estimated yields were folded with the detector response function to compare with the experimental data. An attempt was made to  include the conservation of energy  
by multipliying the bremsstrahlung yield with a factor of $(1-\hbar\omega/{E})$, where ${\hbar\omega}$ 
is the energy carried away by the bremsstrahlung photons. 
The theoretical predictions of the bremsstrahlung yield are shown in Fig.\ref{brem}. 
The dashed line represents the pure Coulomb calculation  i.e without taking into 
account the conservation of energy and pre-scission kinetic energy. The solid line corresponds 
to the calculation performed  considering only conservation of energy while the dot-dashed line 
corresponds to the calculation taking into account both the conservation of energy and pre-scission kinetic energy (x$_{scission}$ = 21.7 fm corresponding to pre-scission kinetic energy of 30 MeV ). As expected, the emission probability is suppressed for higher energies when the 
kinetic energy achieved before reaching scission point is taken into account.

\begin{figure}
\begin{center}
\includegraphics[height=8.5 cm, width=7 cm]{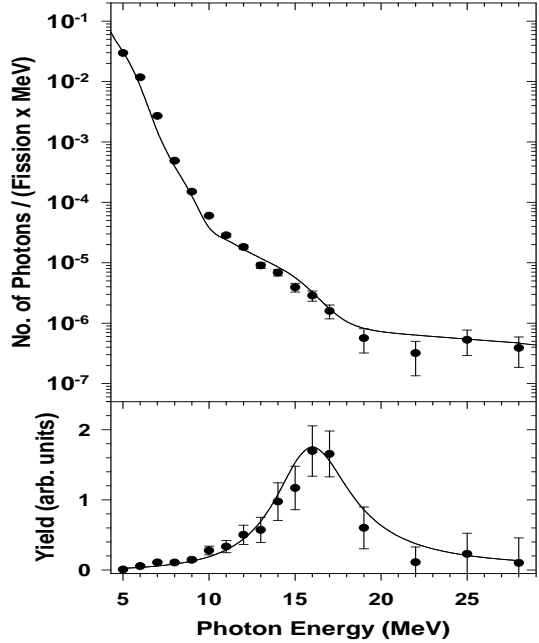}
\caption{\label{gdr}The experimental $\gamma$-spectrum (top) along with the linearized GDR strength function (bottom). The symbol represents the experimental data while the solid line is for CASCADE prediction.}
\end{center}
\end{figure}

In order to calculate the emission of gamma-rays from the decay of GDR accompanying the spontaneous fission of $^{252}$Cf, a modified version of the statistical code CASCADE \cite{cas} was used. Here, only the emission of $\gamma$-rays from the excited fission fragments has been considered, neglecting the pre-scission $\gamma$ contribution. The latter was found to be very small even by increasing the scission time scale \cite{hof93}.
The total $\gamma$-ray spectrum was generated by summing all the gamma spectra calculated independently for all the possible fission fragments and weighed according to  corresponding masses. For each fragment the charge number has been estimated from the relation Z$_{frag}$=A$_{frag}$98/252 \cite{hof93, gup75}. In all the nuclei, Reisdorf-Ignatuyk level density prescription \cite{igna75, reis81} has been used to incorporate the mass and the excitation energy dependence of the level density parameter. The GDR strength function was calculated using a Lorentzian having a centroid energy (E$_{GDR}$) and width ($\Gamma_{GDR}$). The  parameters were calculated dynamically for each fragment mass inside the CASCADE using the systematics  $E_{GDR}$ = 18.0A$^{-1/3}$ + 25.0A$^{-1/6}$ \cite{hara01} and $\Gamma_{GDR}$ = 4.8 + 0.0026E$^{*1.6}$ \cite{chak87}. The high energy photon spectrum, estimated above, was folded with the detector response and is shown in Fig.\ref{gdr}. The bremsstrahlung component has been extrapolated to the lower energies while reconstructing the gamma spectrum. The present calculation represents the experimental data quite well. The linearized GDR lineshape along with the CASCADE prediction is shown in Fig. \ref{gdr}.

\begin{figure}
\begin{center}
\includegraphics[height=7 cm, width=7.5 cm]{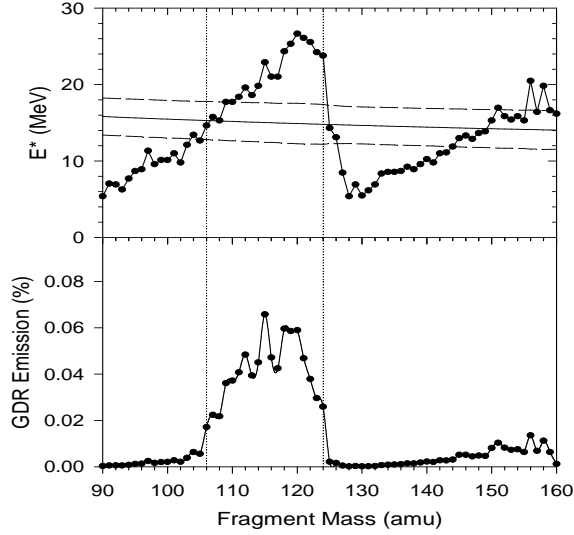}
\caption{\label{mass}(Top)The filled circles represent the excitation energy distribution as a function of fragment mass. The solid line represents the locus of GDR centroid energy while the top and bottom dashed lines correspond to the GDR width about the centroid energy. (Bottom) The filled circles represent the GDR emission probability weighed over corresponding fragment mass. The vertical dotted lines represent the mass region 106-124.}
\end{center}
\end{figure}


\begin{table}
\caption{\label{tab:nucl} Comparision between GDR width measured experimentally and TSFM prediction}
\begin{center}		
\begin{tabular}{|c|c|c|c|}
\hline
Average  & Average    &   Experimental        &    Calculated GDR \\
Mass    & Temp       &     GDR Width         &     Width (TSFM)   \\ \hline
117     &$\;$ 0.68 (MeV)$\;$ &$\;$  5.24 $\pm$ 1.0 (MeV) $\;$&       6.6 (MeV)     \\ 
\hline
\end{tabular}
\end{center}		
\end{table}

In order to understand the emission of GDR photons from the fission fragments, the mass dependent excitation energy was obtained from ref \cite{ploeg95} (top panel of Fig. \ref{mass}). The GDR decay probability was calculated \cite{Lynn} for each mass based on the available excitation energy and is shown in Fig. \ref{mass}(bottom panel) after
weighing over the corresponding mass yield.  The solid line in Fig. \ref{mass}(top panel) corresponds to the locus of E$_{GDR}$ while the dashed lines represent the width of the resonance about centroid energy. It is  evident that the GDR emission probability below this resonance band is negligible. The high energy spectrum contains GDR decay mostly from the fission fragments within the mass region 106-124 without having appreciable contributions from other masses. Interestingly, more than 75 $\%$ of the total GDR decay is from the mass region 109-124 and in addition to that,  E$_{GDR}$ does not change much ($\approx$ 0.4 MeV) for A=109-124. Thus, the measured width actually provides an average width for the mass region 109-124. The temperature was calculated from the initial excitation energy of the fragments after subtracting the rotational energy and the corresponding GDR centroid energy. The average temperature and mass of the region 109-124 were found to be 0.68 MeV and 117 , respectively. These mean values were also estimated by weighing over the GDR emission probability shown in the bottom panel of Fig. \ref{mass}. The extracted GDR width of the average mass $\sim$ 117 is found to be 5.24 $\pm$ 1 MeV. It is observed that the GDR width measured in this experiment is appreciably smaller than the  TSFM predictions (Table 1). The phenomenological description based on the thermal fluctuation theory, describes on the average, many experimental results but it fails to reproduce the data corresponding to the lowest temperatures showing the limitation of the model \cite{kus98}.

In conclusion, the nucleus-nucleus coherent bremsstrahlung from $^{252}$Cf cold fission has been observed and the   
result has been corroborated with a theoretical calculation based on the Coulomb acceleration model. 
The  $^{252}$Cf source provides us an unique tool to study the GDR width at low temperature 
which has been an intriguing topic in the recent years. The GDR widths from the decay of excited fission
fragments have been extracted in order to test the thermal shape fluctuation theory at low temperature. 
The model overpredicts the variation of GDR width at low temperature.   

The authors wish to thank Dr.A.K.Sinha of UGC-DAE CSR for
providing $^{252}$Cf source used in the work.



\end{document}